\documentclass[
secnumarabic, amssymb, amsmath, nofootinbib,tightenlines, showpacs,
nobibnotes, aps, prl]{revtex4}
\usepackage{amsfonts}
\usepackage{docs}%
\usepackage{bm}%
\usepackage[colorlinks=true,linkcolor=blue]{hyperref}%
\usepackage{graphicx}
\usepackage{dcolumn}

\begin{document}

\title{Minimal Conductivity of Topological Surface States with Magnetic Impurities}

\author{Liang Chen}
\author{Shaolong Wan}
\altaffiliation{Corresponding author} \email{slwan@ustc.edu.cn}
\affiliation{Institute for Theoretical Physics and Department of Modern Physics \\
University of Science and Technology of China, Hefei, 230026, {\bf
P. R. China}}

\date{\today}

\begin{abstract}
In this paper we use the semiclassical Boltzmann equation to
investigate the transport properties of Dirac fermion on the surface
of topological insulator with magnetic impurities. The results
obtained show that there is also a minimal conductivity in this
system as in graphene.  We also argue the low temperature transport
property, and find that there is no low temperature anomaly known as
Kondo effect when the temperature is \(T>10^{-6}K\).
\end{abstract}


\maketitle

\section{Introduction}

Recently topological insulator has been theoretically predicted and
experimentally observed in HgTe quantum wells\cite{1,2}, in
\(\mathtt{Bi_{1-x}Sb_{x}}\) alloys\cite{3,4}, and in
\(\mathtt{Bi_2Se_3}\) and \(\mathtt{Bi_2Te_3}\) bulk
crystals\cite{5,6,Chen,Hsieh_3}. The topological insulator is a new
discovered novel material with gapped bulk phase and robust gapless
surface(edge) state. And novel properties of topological insulator
have been predicted, for instance, effective monopole and
topological magnetoelectric effect\cite{Qi_1}, superconductor
proximity effect induced Majorana fermion states\cite{Fu_3}
\textit{etc}. \(\mathtt{Sb_{1-x}Te_x}\) is the first material has
been reported to be topological insulator\cite{4}, and
\(\mathtt{Bi_2Se_3}\), \(\mathtt{Bi_2Te_3}\), \(\mathtt{Sb_2Te_3}\)
have been predicted to be topological insulator\cite{5} with single
Dirac cone on the surface. This material has potential applications
in spintronics, topological quantum computation,
\textit{etc}\cite{Moore_2}.

Recently, nanoribbons of \(\mathtt{Bi_2Se_3}\) have been
fabricated\cite{Peng}, and transport properties of magnetic doped
nanoribbons have been measured\cite{Cha} with conductivity Kondo
effect. However, the experiment can't distinct the conductivity
anomaly from Kondo effect induced by bulk defect conductance. In
this paper, we use the semi-classical Boltzmann equation to
investigated the transport properties of topological insulator
surface state with surface magnetic impurity doped.

The paper is organized as follows: In Sec. II, we construct an
effective interaction between topological insulator surface state
and magnetic impurities, and give the corresponding Boltzmann
equation with coherent distribution function. In Sec. III, we
analyze the incoherent case and give an exact solution for the
Boltzmann equation. We also take into account the conductivity
correction from electron-hole coherent terms in Sec. IV. In Sec. V,
we consider the high-order correction of coupling constant, and give
the renormalization group equation of the the interaction between
TISS and magnetic impurities. We make a conclusion in Sec. VI.

\section{Hamiltonian and Boltzmann Equation}
\indent At first we take \(\mathtt{Sb_2Te_3}\) as an example to
construct the Hamiltonian and make some assumptions: (i) The
chemical potential has been tune to seat at the Dirac point, which
has been realized by D.Hsieh \textit{et al.}\cite{Hsieh_2} (ii) The
interaction between topological insulator surface state(TISS) and
magnetic impurities take the form of spin-spin interaction. (iii)
The effective couplings are short-range rotational symmetric
potential, which can be simulated by \(J_{\mu}(\mathbf{r}) =
J_{\mu}e^{-r/R}/r\),
where exchange parameters \(J_{\mu}\) are estimated to the order
\(0.1\text{eV}\sim0.5\text{eV}\)\cite{QLiu,Dyck}, and the range R of
interaction has been assume to about 13{\AA}\cite{QLiu}. (iv) Due to
the effective RKKY interaction between magnetic impurities, all of
the impurities have been ranged in the same direction\cite{QLiu}.
(v)The concentration of magnetic impurity is very small, so we can
take the single impurity approximation. According to above, the
Hamiltonian of this system can be described by:
\begin{equation}
H_0 = \hbar v_F\int \mathtt{d}^2 r C^{\dagger}(\mathbf{r}) i
\sigma^{\alpha} \partial_{\alpha} C(\mathbf{r}), \label{2.1}
\end{equation}
\begin{equation}
H_{int} = \int \mathtt{d}^2 \mathbf{r}
J_{\mu}(\mathbf{r})\hat{S^{\mu}} C^{\dagger}(\mathbf{r})
\hat{\sigma}^{\mu} C(\mathbf{r}), \label{2.2}
\end{equation}
where \(C(\mathbf{r}), C^{\dagger}(\mathbf{r})\) are annihilation, creation operators of TISS respectively, \(\sigma^{\alpha}\) \((\alpha = x,y)\) are the Pauli matrices in spin space of TISS, \(\hat{S}^{\mu}\) \((\mu = x,y,z)\) is the spin operators of impurities and \(\hbar v_F \simeq \text{3.7eV}\cdot\){\AA}(\(v_F\) is Fermi velocity). The eigenstates of free Hamiltonian take:
\begin{equation}
|\psi_{+}(\mathbf{k})\rangle = \frac{1}{\sqrt{2}}\left(\begin{array}{c}
                                                  e^{-i\theta} \\
                                                  1
                                                \end{array}\right) , \ \
|\psi_{-}(\mathbf{k})\rangle = \frac{1}{\sqrt{2}}\left(\begin{array}{c}
                                                  e^{-i\theta} \\
                                                  -1
                                                \end{array}\right), \label{2.3}
\end{equation}
with corresponding eigenvalues \(\epsilon_{\pm}(\mathbf{k}) = \pm\hbar v_F k\) and \(\theta\) is the azimuth angle \(\arctan k_y / k_x\). Under these basis, the scattering of TISS by magnetic impurity can be expressed as:
\begin{eqnarray}
\label{Eq_int}
H_{int}(\mathbf{k_1},\mathbf{k_2}) &=& \frac{1}{2}\sum_{\mu, \nu = \pm1} \left[J_z(e^{i(\theta_1-\theta_2)}-\mu\nu) \hat{S}^z + \nu J_{\shortparallel} e^{i\theta_1}\hat{S}^- + \mu J_{\shortparallel} e^{-i\theta_2}\hat{S}^+\right] \nonumber\\
& & C^{\dagger}_{\mu}(\mathbf{k_1}) C_{\nu}(\mathbf{k_2}),
\label{2.4}
\end{eqnarray}
here \(J_{\mu}\) is the short-written of \(J_{\mu}\left(|\mathbf{k_1}-\mathbf{k_2}|^2 + \frac{1}{R^2}\right)^{-1/2}\), and we assume \(J_x = J_y = J_{\shortparallel}\). \(\theta_1\), \(\theta_2\) are the azimuth angle of \(\mathbf{k_1}\) and \(\mathbf{k_2}\).\\
Electrical current of this system reads:
\begin{equation}
\label{Eq_J} \mathbf{J} = \int
\frac{\mathtt{d}^2k}{(2\pi)^2}\,tr\left[e\hat{\mathbf{v}}\hat{f}(\mathbf{k})\right],
\end{equation}
where \(\hat{f}(\mathbf{k})\) is particle number distribution, and the velocity operator \(\hat{\mathbf{v}}\) is a \(2 \times 2\) matrix in the space expand by \(\{\psi_+, \psi_-\}\):
\begin{equation}
\frac{\hat{\mathbf{v}}}{v_F} = \left(
                                 \begin{array}{cc}
                                   \mathbf{e}_x\cos\theta + \mathbf{e}_y\sin\theta & -i(\mathbf{e}_x\sin\theta - \mathbf{e}_y\cos\theta) \\
                                    i(\mathbf{e}_x\sin\theta - \mathbf{e}_y\cos\theta) & -\mathbf{e}_x\cos\theta - \mathbf{e}_y\sin\theta \\
                                 \end{array}
                               \right), \label{2.6}
\end{equation}
here \(\theta\) is  the azimuth angle of \(\mathbf{v}\).\\
The Boltzmann equation, up to linear order of homogeneous electric field, of a similar system in graphene has been derived in \textit{reference} \cite{Trushin}:
\begin{eqnarray}
\label{Eq_Bol}
\left(\frac{\mathtt{d}\hat{f}}{\mathtt{d}t}\right)^{coll} &=& \frac{i}{\hbar}
\left(
  \begin{array}{cc}
    0 & f^{+-}(\epsilon_+(\mathbf{k}) - \epsilon_-(\mathbf{k})) \\
    f^{-+}(\epsilon_-(\mathbf{k}) - \epsilon_+(\mathbf{k})) & 0 \\
  \end{array}
\right) + \nonumber\\
& & \frac{e}{\hbar}\mathbf{E}\cdot\left(
                                    \begin{array}{cc}
                                      \hbar\mathbf{v_{11}}\frac{\partial f^{++}_{0}(\epsilon_+(\mathbf{k}))}{\partial\epsilon_+(\mathbf{k})} & \frac{1}{2}\frac{\hbar\mathbf{v_{12}}}{\epsilon_+(\mathbf{k})} \left(f^{++}_0- f^{--}_0\right)\\
                                      \frac{1}{2}\frac{\hbar\mathbf{v_{21}}}{\epsilon_-(\mathbf{k})} \left(f^{--}_0- f^{++}_0\right) & \hbar\mathbf{v_{22}}\frac{\partial f^{--}_{0}(\epsilon_-(\mathbf{k}))}{\partial\epsilon_-(\mathbf{k})} \\
                                    \end{array}
                                  \right), \label{2.7}
\end{eqnarray}
and the elastic collision term can be expressed by\cite{Dyakonov}:
\begin{eqnarray}
\label{Eq_coll}
\left(\frac{\mathtt{d}\hat{f}}{\mathtt{d}t}\right)^{coll}_{\mu\nu} &=& \frac{1}{2}\sum_{\alpha\beta=\pm} \int \frac{\mathtt{d}^2k'}{(2\pi)^2} \left\{\left[\delta(\epsilon_{\mu}-\epsilon_{\alpha}) + \delta(\epsilon_{\mu}-\epsilon_{\beta})\right] K^{\mu\nu}_{\alpha\beta}f_{\alpha\beta}(\mathbf{k'}) \right.  \nonumber\\
 & &  \;\;\left. -\delta(\epsilon_{\alpha}-\epsilon_{\beta})\left[K^{\mu\alpha}_{\beta\beta}f_{\alpha\nu}(\mathbf{k}) + K^{\alpha\nu}_{\beta\beta}f_{\mu\alpha}(\mathbf{k})\right]\right\}.
 \label{2.8}
\end{eqnarray}
where \(f^{\mu\nu}\) (\(\mu,\nu=\pm\)) are the four components of
\(\hat{f}(\mathbf{k})\), \(f_0^{\mu\nu}\) are equilibrium
distribution function without external electrical field, E is the
homogeneous external electrical field, and \(\mathbf{v}_{ij}\)
(i,j=1,2) are the four elements of velocity matrix.

\section{Incoherent Case}

In equation (\ref{Eq_Bol}) and (\ref{Eq_coll}), by setting the
coherent terms \(f^{-+}(\mathbf{k}), f^{+-}(\mathbf{k})\) to be
zero, we can get a more familiar Boltzmann transport equation of the
form:
\begin{equation}
\label{Eq_transpt} e\mathbf{E}\cdot\mathbf{v}_{\mu} = \int
\frac{\mathtt{d}^2k'}{(2\pi)^2}\left(\frac{2\pi}{\hbar}\right)
\delta(\epsilon_{\mu}(\mathbf{k})-\epsilon_{\mu}(\mathbf{k'}))
|\langle\psi_{\mu}(\mathbf{k})|\hat{T}|
\psi_{\mu}(\mathbf{k'})\rangle|^2\left(f_\mu(\mathbf{k'}) -
f_{\mu}(\mathbf{k})\right),
\end{equation}
where \(\mu=\pm\) for \(\psi_{\pm}(\mathbf{k})\) states, \(\hat{T}\) is the transition matrix, \(\mathbf{E}\) is the homogeneous external electric field. Under the first order of coupling constant \(\hat{T}\)-Matrix can be expressed as:
\begin {equation}
\label{Eq_Tmatx} \langle u(\mathbf{k_1})
|\hat{T}|u(\mathbf{k_2})\rangle = \frac{1}{2} \left[ J_z
(e^{i(\theta_1-\theta_2)}-1)\hat{S}_z +
J_{\shortparallel}e^{i\theta_1}\hat{S}^- +
J_{\shortparallel}e^{-i\theta_2}\hat{S}^+ \right],
\end{equation}
and under the relaxation time approximation up to linear order of external homogeneous electric field, we have
\begin{equation}
\label{Eq_relax} f_{\mu}(\mathbf{k}) = f^{0}_{\mu}(\mathbf{k}) +
e\mathbf{E}\cdot\mathbf{v}_{\mu} \tau_{\mu}(\mathbf{k})
\frac{\partial f^{0}_{\mu}(\mathbf{k})}{\partial
\epsilon_{\mu}(\mathbf{k})}.
\end{equation}
According to equation (\ref{Eq_transpt}), (\ref{Eq_Tmatx}) and (\ref{Eq_relax}), we can find an exact analytical solution for relaxation time:
\begin{eqnarray}
\frac{1}{\tau_{\pm}(\mathbf{k})} &=& \frac{kR^2N}{2 \hbar^2 v_F \sqrt{1+4R^2k^2}} \nonumber\\
 & & \left\{
J_z^2 S^z S^z\left[1 - \left(3+\frac{1}{2R^2k^2}\right)\frac{\sqrt{1+4R^2k^2}-(1+2R^2k^2)}{2R^2k^2}\right]\right.   \nonumber\\
 & & \left. + J_{\shortparallel}^2(S(S+1)-S^zS^z)\left[1-\frac{\sqrt{1+4R^2k^2}-(1+2R^2k^2)}{2R^2k^2}\right]
\right\}, \label{3.4}
\end{eqnarray}
here N is the concentration of magnetic impurities on topological insulator surface. In the short-range limit \(R k \ll  1\), relaxation time takes:
\begin{equation}
\tau_{\pm}(\mathbf{k}) = \frac{1}{k} \frac{2\hbar^2 v_F}{R^2N}
\frac{1}{\frac{3}{2}J_z^2S^zS^z +
J_{\shortparallel}^2(S(S+1)-S^zS^z)}. \label{3.5}
\end{equation}

\section{Coherent Terms Included}

In this section, we take into account contributions of coherent
terms in equation (\ref{Eq_Bol}) and (\ref{Eq_coll}), the
non-equilibrium part of the contribution function in these equations
can be written as:
\begin{eqnarray}
\label{Eq_expad}
f^{1}_{++}(\mathbf{k}) &=& e E v_F\left[\cos(\theta-\phi)\tau_{11}(k) + \sin(\theta-\phi)\lambda_{11}(k)\right] , \nonumber\\
f^{1}_{+-}(\mathbf{k}) &=& e E v_F\left[- i \sin(\theta-\phi)\tau_{12}(k) - i \cos(\theta-\phi)\lambda_{12}(k)\right] , \nonumber\\
f^{1}_{-+}(\mathbf{k}) &=& e E v_F\left[ i \sin(\theta-\phi)\tau_{21}(k) + i \cos(\theta-\phi)\lambda_{21}(k)\right] , \nonumber\\
f^{1}_{--}(\mathbf{k}) &=& e E
v_F\left[-\cos(\theta-\phi)\tau_{22}(k) -
\sin(\theta-\phi)\lambda_{22}(k)\right],
\end{eqnarray}
here \(\theta\) is the azimuth angle \(\arctan(k_y/k_x)\) and \(\phi\) determines the direction of electric field \(\mathbf{E} = \text{E}(\cos\phi,\,\sin\phi)\). It can be verified that longitudinal current is proportional to the terms contain \(\tau_{ij}(k)\) and transverse current is proportional to the terms contain \(\lambda_{ij}(k)\). And we will show that the transverse current induced by magnetic impurities vanishes. \\

We calculated all of the matrix elements \(K^{\alpha\beta}_{\gamma\delta}\):
\begin{equation}
K^{\alpha\beta}_{\gamma\delta} = \frac{\pi}{\hbar} (\langle
\psi_{\alpha}(\mathbf{k})|\hat{T}|\psi_{\gamma}(\mathbf{k'})\rangle)
(\langle
\psi_{\beta}(\mathbf{k})|\hat{T}|\psi_{\delta}(\mathbf{k'})\rangle)^{*},
\nonumber \label{4.2}
\end{equation}
replaced non-equilibrium terms of \(f_{\mu\nu}(\mathbf{k})\) with equation (\ref{Eq_expad}), and took some tedious calculations from equations(\ref{Eq_Bol}) and (\ref{Eq_coll}), then we found that \(\tau_{ij}(k)\) and \(\lambda_{ij}(k)\) satisfy the following equations (in the short-range limit \(R k \ll 1\)):
\begin{equation}
  \left\{
     \begin{aligned}
        -2(\alpha + \xi ) \tau_{11} &- \alpha(\tau{12}+\tau_{21}) + i \zeta (\lambda_{12}-\lambda_{21}) = \frac{\hbar^2v_F}{k} \frac{\partial f^{0}_{11}}{\partial \epsilon_{1}}
        \\
        -2(\alpha + \xi ) \lambda_{11} &+ \alpha(\lambda_{12}+\lambda_{21}) + i \zeta (\tau_{12}-\tau_{21}) =0
        \\
        -\zeta(\tau_{11}-\tau_{22}) &- i \alpha (\lambda_{11}+\lambda_{22}-\lambda_{21}) + i (2\xi +\alpha) \lambda_{12} = \frac{\hbar^2v_F}{k}\frac{\epsilon_1 - \epsilon_2}{\hbar}\lambda_{12}
        \\
        -\zeta (\lambda_{11} - \lambda_{22}) &+ i \alpha (\tau_{11}+\tau_{22}+\tau_{21}) + i(2\xi+\alpha) \tau_{12} = \frac{\hbar^2v_F}{k}\left(\frac{\epsilon_1 - \epsilon_2}{\hbar}\tau_{12} - \frac{i}{2}\frac{f^{0}_{11}-f^{0}_{22}}{\epsilon_1}\right)
        \\
        (1 \longleftrightarrow  2) &
        \\
     \end{aligned}
  \right. , \label{4.3}
\end{equation}
where \(\alpha\), \(\xi\) and \(\zeta\) are three parameters: \(\alpha = \frac{N}{8}R^2J_z^2S^zS^z\), \(\xi = \frac{N}{8}R^2[2J_z^2S^zS^z + 2J_{\shortparallel}^2(S(S+1)-S^zS^z)]\), \(\zeta = -\frac{N}{8}R^2 J_{\shortparallel}^2 S^z\). \((1 \leftrightarrow 2) \) means exchange indexes 1 and 2 to get another four equations. By solving Eq.(15), we find a solution to equations (\ref{Eq_Bol}) and (\ref{Eq_coll}) in the short-range limit, which are:
\begin{eqnarray}
f^{1}_{11}(\mathbf{k}) &=& e E v_F\cdot \tau(k)\left\{
    \cos(\theta-\phi)
        \left[(1+\frac{1}{F})\left(-\frac{\partial f^{0}_{11}}{\partial \epsilon_1}\right) + \frac{\alpha}{\alpha+\xi}\frac{1}{F} \left(\frac{f^{0}_{11}-f^{0}_{22}}{2\epsilon_1}\right) \right] \right.\nonumber\\
    & &\left.-\sin(\theta-\phi)\frac{1}{H}\left[
        \frac{\alpha}{\alpha+\xi}\left(-\frac{\partial f^{0}_{11}}{\partial \epsilon_1}\right) + \left(\frac{f^{0}_{11}-f^{0}_{22}}{2\epsilon_1}\right)
        \right]
       \right\},  \nonumber\\
f^{1}_{22}(\mathbf{k}) &=& -e E v_F\cdot \tau(k)\left\{
    \cos(\theta-\phi)
        \left[(1+\frac{1}{F})\left(-\frac{\partial f^{0}_{22}}{\partial \epsilon_2}\right) + \frac{\alpha}{\alpha+\xi}\frac{1}{F}\left(\frac{f^{0}_{22}-f^{0}_{11}}{2\epsilon_2}\right) \right] \right.\nonumber\\
    & &\left.+\sin(\theta-\phi)\frac{1}{H}\left[
        \frac{\alpha}{\alpha+\xi}\left(-\frac{\partial f^{0}_{22}}{\partial \epsilon_2}\right) + \left(\frac{f^{0}_{22}-f^{0}_{11}}{2\epsilon_2}\right)
        \right]
       \right\},  \nonumber\\
f^{1}_{12}(\mathbf{k}) &=& i e E v_F\cdot \tau(\mathbf{k})\sin(\theta-\phi)\left(\frac{1}{F} - i\frac{1}{G}\right)\left[\frac{\alpha}{\alpha+\xi}\left(-\frac{\partial f^{0}_{11}}{\partial \epsilon_1}\right) + \left(\frac{f^{0}_{11}-f^{0}_{22}}{2\epsilon_1}\right)\right], \nonumber\\
f^{1}_{21}(\mathbf{k}) &=& -i e E v_F\cdot
\tau(\mathbf{k})\sin(\theta-\phi)\left(\frac{1}{F} +
i\frac{1}{G}\right)\left[\frac{\alpha}{\alpha+\xi}\left(-\frac{\partial
f^{0}_{22}}{\partial \epsilon_2}\right) +
\left(\frac{f^{0}_{22}-f^{0}_{11}}{2\epsilon_2}\right)\right],
\nonumber\\ \label{4.4}
\end{eqnarray}
here we define:
\begin{equation*}
\frac{1}{F} = \frac{\xi(\alpha+\xi)-\zeta^2}{\hbar^4 v_F^4}, \;\;
\frac{1}{G} = \frac{\alpha+\xi}{\hbar^2 v_F^2}, \;\; \frac{1}{H} =
\frac{\zeta}{\hbar^2 v_F^2}. \label{4.5}
\end{equation*}
And the new relaxation time with coherent term correction is determined by
\begin{equation}
\label{Eq_TE} \tau(k)\cdot|\epsilon(k)| = \frac{\hbar
G/2}{1+\frac{\xi(2\alpha+\xi)}{(\alpha+\xi)^2}\frac{1}{F}}.
\end{equation}
The conductivity of this system can be deduced from equation (\ref{Eq_J}) straightforwardly, while the longitudinal conductivity reads:
\begin{equation}
\sigma = \frac{e^2 v_F^2}{2\pi} \int_0^\infty k \mathtt{d}k \,
\tau(k)\left[-\frac{\partial f^{0}_{11}}{\partial \epsilon_1} +
\frac{1}{F}\left(\frac{2\alpha+\xi}{\alpha+\xi}\right)\left(-\frac{\partial
f^{0}_{11}}{\partial \epsilon_1} +
\frac{f^{0}_{11}-f^{0}_{22}}{2\epsilon_1}\right)\right], \label{4.7}
\end{equation}
and the contribution to transverse current from terms in \(f^{1}_{11}(\mathbf{k})\) and \(f^{1}_{22}(\mathbf{k})\) proportional to \(1/H\) have been canceled to each other.

\section{High Order Corrections}

Assuming the correction from coherent terms is small \(\frac{1}{F}
\ll 1\), so the conductivity correction from high order terms of
transition matrix may be taken into considered only in the
incoherent case. In the short-range regime, Poorman's
renormalization suggest that traditional RG equation\cite{Hewson} is
still correct in this system:
\begin{equation}
\label{Eq_Kondo}
    \left\{
        \begin{aligned}
            \frac{\mathtt{d}(RJ_z)}{\mathtt{d}l} &= -\frac{1}{2\pi\hbar^2v_F^2}\frac{D}{1+\left(\frac{RD}{\hbar v_F}\right)^2} (RJ_{\shortparallel})^2
            \\
            \frac{\mathtt{d}(RJ_{\shortparallel})}{\mathtt{d}l} &= -\frac{1}{2\pi\hbar^2v_F^2}\frac{D}{1+\left(\frac{RD}{\hbar v_F}\right)^2} RJ_{\shortparallel}\cdot RJ_z
        \end{aligned}
    \right.,
\end{equation}
where \(D\) is the high energy truncation, with the magnitude of bulk energy gap, \(l\) is the renormalization rescaling factor: \(l = \log(D/T)\). Higher order correction of transition matrix elements have the form:
\begin{equation}
\begin{split}
&\langle \psi_{\mu}(\mathbf{k_1})|\hat{T}|\psi_{\mu}(\mathbf{k_2})\rangle^{(2)}  =
\\
&\sum_{\mathbf{k},\rho=\pm}\frac{1/4}{\epsilon_{\mu}(\mathbf{k_2}) - \epsilon_{\rho}(\mathbf{k})}
\left[
\left( J_0^2 + J_z^2S^zS^z+J_{\shortparallel}^2S(S+1)-J_{\shortparallel}^2S^zS^z\right)
\left(e^{i(\theta_1-\theta_2)} + 1 \right)\right.
\\
& \left.+ 2J_0J_z\left(e^{i(\theta_1-\theta_2)} - 1 \right) S^z
  + 2\mu J_0J_{\shortparallel}e^{i\theta_1}S^{-}
 + 2\mu J_0J_{\shortparallel}e^{-i\theta_2}S^{+}
    \right]
\\
&+ \sum_{\mathbf{k},\rho=\pm}\frac{f_{\rho}(\mathbf{k})/2}{\epsilon_{\mu}(\mathbf{k_2}) - \epsilon_{\rho}(\mathbf{k})}
 \left[
    J_{\shortparallel}^2\left(e^{i(\theta_1-\theta_2)} - 1 \right) S^z
 +\mu J_zJ_{\shortparallel}e^{i\theta_1}S^{-}
 +\mu J_zJ_{\shortparallel}e^{-i\theta_2}S^{+}
\right],
\end{split} \label{5.2}
\end{equation}
here we have considered an additional screened Coulomb potential in the interaction Hamiltonian, which has been induced by higher order interactions of TISS and magnetic impurities, so \(J_0 \ll J_{z,\shortparallel}\). It can be proved by following the Poorman's renormalization that the RG equation is \(\mathtt{d}J_0/\mathtt{d}l = 0\). Finally the relaxation time correction can be calculated by the following equation:
\begin{equation}
\frac{1}{\tau_{\mu}(\mathbf{k})} = \frac{2\pi N}{\hbar^2v_F} \int
\frac{\mathtt{d}^2 k'}{(2\pi)^2} \delta(k-k') \arrowvert
\langle\psi_{\mu}(\mathbf{k})|\hat{T}|\psi_{\mu}(\mathbf{k'})\rangle
\arrowvert^2 (1-\cos(\theta-\theta')), \label{5.3}
\end{equation}
which reads:
\begin{equation}
\delta\left[\frac{1}{\tau_{\mu}(\mathbf{k})}\right] =
\frac{kN}{2\hbar^2v_F}R^3J_zJ_{\shortparallel}^2S(S+1)\sum_{\mathbf{k'},\rho}\frac{f_{\rho}(k')}
{\epsilon_{\mu}(k)-\epsilon_{\rho}(k')}. \label{5.4}
\end{equation}
And the corresponding conductivity correction is:
\begin{eqnarray}
\delta\sigma &=& -\frac{e^2\hbar^2v_F^3}{NR\pi^3}\frac{J_z J_{\shortparallel}^2S(S+1)}{[3J_z^2 S^zS^z + 2J_{\shortparallel}^2(S(S+1)-S^zS^z)]^2}\times  \nonumber\\
& &\int_0^\infty \mathtt{d}k \left(-\frac{\partial
f^0_{+}(k)}{\partial \epsilon_+(k)}\right)\int \mathtt{d}^2k'
\sum_{\mu,\rho=\pm}\frac{f_\rho(k')}{\epsilon_{\mu}(k)-\epsilon_{\rho}(k')}.
\label{5.5}
\end{eqnarray}
The integration over \(\mathbf{k'}\) has been limited in a finite regime, because these conductivity correction is induced by a virtual state \(|\psi_{\rho}(k')\rangle\) while \(|\psi_{\mu}(k')\rangle\) was propagating, and the uncertainty principle needs \(|\epsilon_{\rho}(k')-\epsilon_{\mu}(k)|\tau(k')\le\hbar\). Then according to equation (\ref{Eq_TE}), the boundary of \(\epsilon_{\rho}(k')\) takes the form \(\epsilon_{\rho}^{c}(k') = \epsilon_{\mu}(k)/(1\pm\frac{2}{G})\). The conductivity correction can be rewritten as:
\begin{eqnarray}
\delta\sigma &=& -\frac{e^2}{h}\frac{T{NR}}{4\pi\hbar v_F}\frac{R^3 J_zJ_{\shortparallel}^2S(S+1)}{(\hbar v_F)^3}\left(\frac{2}{G}\right)^{-2}\kappa(2/G),
\\
\kappa(2/G) &=& \int_0^{\infty} \mathtt{d}x \frac{xe^x}{(e^x+1)^2}
\int_{-2/G}^{2/G}
\mathtt{d}\lambda\left(1+\frac{1}{\lambda}\right)\frac{1-e^{-x(1+\lambda)}}{1+e^{-x(1+\lambda)}}.
\nonumber \label{5.6}
\end{eqnarray}
where \(\kappa(2/G)\) (\(0<2/G<1\)) is a nondimensional function of
\(2/G\). So the high-order conductivity correction \(\delta\sigma\)
proportional to temperature. In the low temperature limit, coupling
constants flow into a strong coupled regime
\(2/G\rightarrow\infty\), this semi-classical Boltzmann equation
method is invalid because equation (\ref{Eq_TE}) is not consistent
with uncertainty principle. For this reason, it is necessary to
discuss the validity of our results obtained furthermore. According
to equation(\ref{Eq_Kondo}), we estimate the Kondo temperature of
this system and obtain \(T_K = D \exp\{-\frac{4\pi(\hbar
v_F)^2}{R^2DJ}\}\), where D take the value of bulk energy gap which
is about 0.1eV for \(\mathtt{Sb_2Te_3}\),  R is about
\(13\)\AA\cite{QLiu}, and antiferromegnetic coupling is assumed to
be isotropy with value 3.5eV. We get Kondo temperature is about
\(10^{-6}\)K. So we obtain that surface magnetic dropped
\(\mathtt{Sb_2Te_3}\) doesn't reveal low temperature conductivity
anomaly when temperature is not very low and hope be observed in
future.

\section {Conclusion}

In this paper we investigate the transport properties of Dirac
fermion on the surface of topological insulator with magnetic
impurities. We find that there is also a minimal conductivity in the
system studied as in graphene and there is no low temperature
anomaly known as Kondo effect when the temperature \(T>10^{-6}K\).

\section*{Acknowledgement}

This work is supported by NSFC Grant No.10675108.

\end{document}